# Exploring the Protein Sequence Space with Global Generative Models

Sergio Romero-Romero[1†], Sebastian Lindner[2†], Noelia Ferruz[3*]

[1] Department of Biochemistry, University of Bayreuth, 95447 Bayreuth, Germany
[2] University of Heidelberg, 69047 Heidelberg, Germany
[3] Molecular Biology Institute of Barcelona, 08028 Barcelona, Spain

† These authors contributed equally
* Corresponding author: Noelia Ferruz. E-mail: noelia.ferruz@ibmb.csic.es

## ORCID identifiers and emails

Sergio Romero-Romero: 0000-0003-2144-7912. sergio.romero-romero@uni-bayreuth.de
Sebastian Lindner: 0009-0009-8251-2758. sebastian.lindner@stud.uni-heidelberg.de
Noelia Ferruz: 0000-0003-4172-8201. noelia.ferruz@ibmb.csic.es

## Short title

Generative models across the sequence space

## Abstract

Recent advancements in specialized large-scale architectures for training image and language have profoundly impacted the field of computer vision and natural language processing (NLP). Language models, such as the recent ChatGPT and GPT4 have demonstrated exceptional capabilities in processing, translating, and generating human languages. These breakthroughs have also been reflected in protein research, leading to the rapid development of numerous new methods in a short time, with unprecedented performance. Language models, in particular, have seen widespread use in protein research, as they have been utilized to embed proteins, generate novel ones, and predict tertiary structures. In this book chapter, we provide an overview of the use of protein generative models, reviewing 1) language models for the design of novel artificial proteins, 2) works that use non-Transformer architectures, and 3) applications in directed evolution approaches.

## Introduction

Proteins are highly attractive nanomaterials, capable of performing a wide range of functions under mild, non-toxic conditions. This has prompted significant research efforts in the field of protein design, with a particular emphasis on the development of functional proteins. In the last two decades, remarkable advances have been made in this area, including the design of novel *de novo* protein structures using traditional methods (Korendovych and DeGrado 2020; Romero-Romero et al. 2021; Pan and Kortemme 2021). The conventional approach to protein design involves providing a backbone scaffold as input and then utilizing computational methods to identify optimal sequences that fold into the scaffold. This problem, often referred to as the *inverse folding problem*, has been mathematically formulated as an optimization problem, where the goal is to find the global minimum of a high-dimensional physicochemical energy function (**Fig. 1a**). However, due to its computational complexity - with over $100^{20}$ possible sequences for a protein of 100 amino acids- approximations to both the algorithm and function are often employed (Huang et al. 2016), with the exception of a few algorithms that sample the energy space exhaustively and deterministically (Gainza et al. 2016).

Despite significant progress in the field of protein design in recent years (as detailed in Huang et al. 2016 and Pan and Kortemme 2021), traditional approaches have two primary limitations. Firstly, they typically require a pre-defined protein backbone as input, which may not be the optimal scaffold for a given function. Secondly, the integration of functional properties is typically performed in a separate step after protein design, a process that can be time-consuming, extending over several years. However, the integration of artificial intelligence (AI) methods in protein design has brought about a paradigm shift in the field. The rapid advancement of deep learning (DL) architectures and hardware has led to significant breakthroughs in various fields, such as the development of tools for image processing (DALLE-2, StableDiffusion), text generation (ChatGPT, GPT-3), audio assistants (Siri, Alexa), and even puts self-driving cars in the horizon. In the field of protein research, one notable example of this revolution is the structure prediction method Alphafold2, which has inspired the development of numerous DL-based protein design methods. In fact, more than 40 different methods have been developed in the past three years alone (Ferruz et al. 2022a). These methods provide not only new approaches to the traditional inverse folding problem (**Fig. 1b**) but also introduce novel ways of designing proteins, such as generating structures (Anand and Huang 2018) (**Fig. 1c**), sequences (Nijkamp et al. 2022; Ferruz et al. 2022b) (**Fig. 1d**), or concurrently designing both (Wang et al. 2022) (**Fig. 1e**).

In this chapter, we delve into the potential of models that are capable of generating protein sequences across the entire protein space. We begin by reviewing Transformer-based language models that can generate unconditioned protein sequences or sequences conditioned on a user-defined prompt. Then, we focus on generative models that employ deep learning architectures other than Transformers, such as Generative Adversarial Networks (GANs), Variational Autoencoders (VAEs), or Long-Short Term Memory (LSTM) networks. Finally, we examine the use of DL methods in the field of directed evolution and protein design. Our goal is to provide a comprehensive overview of the use of AI-based methods for sequence generation and to introduce readers to this emerging field of research.

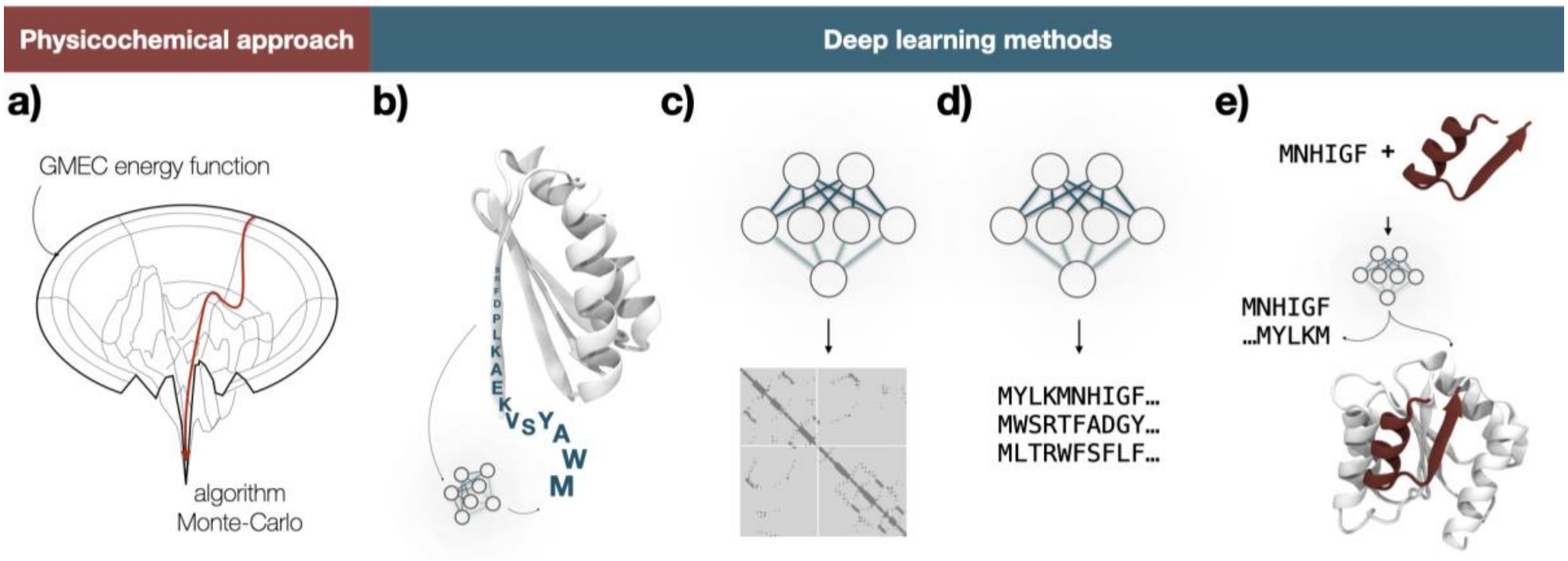

**Figure 1. A paradigm shift in protein design. a)** traditional protein design problem, where an approximated energy function, such as the Global Minimum Energy Conformation (GMEC) is searched with a heuristic algorithm. Deep learning techniques have enabled the design of fixed-backbone sequences **(a)**, structure generation **(b)**, sequence generation **(c),** and sequence and structure design around a scaffold **(d, e).**

## 1. Transformer-based language models

The Transformer has emerged as the most critical development in AI in the last years (Vaswani et al. 2017), enabling the implementation of a myriad of language models. Its success is mainly attributable to the attention mechanism (Bahdanau et al. 2014), which originated as a solution to traditional sequence-to-sequence (seq2seq) models (**Fig. 2**). In seq2seq models, the input (a sentence) is stepwise processed in the encoder to produce a context vector passed to the decoder, an architecture that however exhibited degrading performance and increasing times with sequence length. The attention mechanism provided a solution to these problems since it allows the decoder to analyze the whole input and focus on specific parts, a notion similar to attention in the human mind. A simplified example of the attention mechanism is to focus on the input word '*home*,' when outputting the word '*maison*' in an English-to-French translation (**Fig. 2a**). The Transformer not only mediated the attention mechanism between the two modules but also throughout them, producing a much better performance in many tasks. Following these advances, researchers soon started exploring the modules' performance separately.

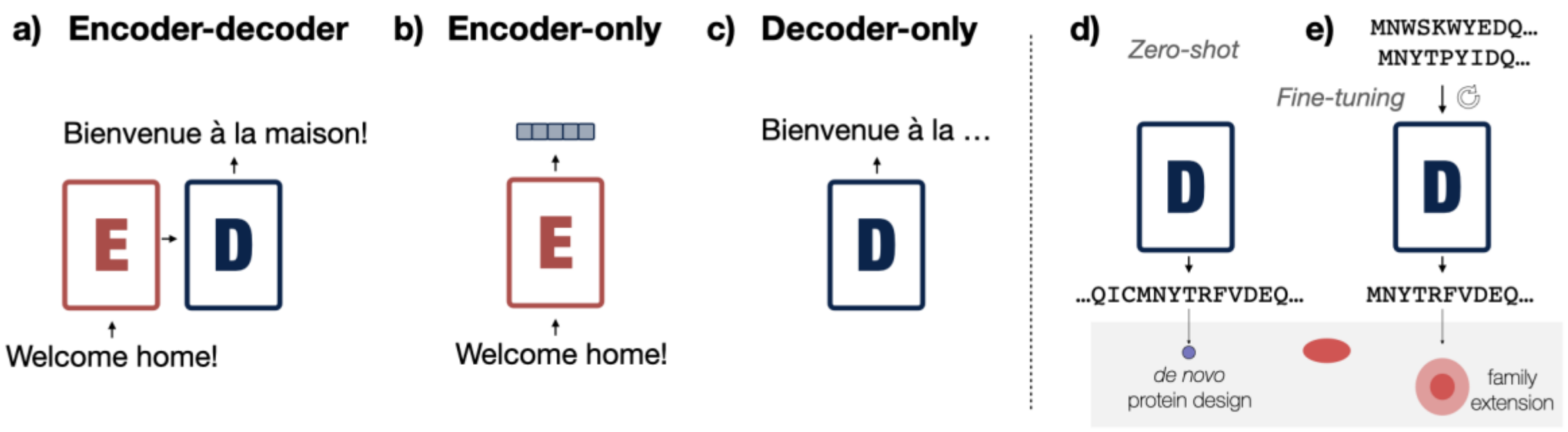

**Figure 2**. **Architectural types of language models and their examples in NLP. a)** The original Transformer has encoder (E) and decoder (D) modules. **b)** The BERT model is based on the original Transformer but contains only the encoder model. **c)** The GPT-n models are based on the decoder-only part of the Transformer and have generative capabilities. **d)** In *zero-shot*, language models generate sequences unconditionally. **e)** During *fine-tuning*, the model updates its weights and can sample new sequences with properties from the new training set.

In this direction, Devlin et al. pre-trained Bidirectional Encoder Representations from Transformers (BERT) (Devlin et al. 2018). BERT is also inspired by the Transformer architecture. Still, given that in this case, the interest lies in creating representations of text input, it only uses the encoder module (**Fig. 2b**). Models like BERT are trained by corrupting the input tokens in some way, e.g., by masking and trying to reconstruct the original sentence, such as in cloze tests. Soon after, OpenAI released GPT (Generative Pretrained Transformer), the first of a series of highly performing generative models, the most recent being ChatGPT and GPT4. GPT was pre-trained on the classic language modeling task, namely, predicting the next item of a sequence based on the previous ones – a task that makes it particularly powerful for language generation. Models trained on this objective are termed *autoregressive,* and their architecture corresponds to the decoder module (**Fig. 2c**).

Given this unprecedented success, researchers soon began to apply BERT and GPT-n architectures to the protein realm. Some of the earliest examples of encoder-only protein language models are ESM-1b (Rives et al. 2021) and ProtTrans (Elnaggar et al. 2021), which showed remarkable accuracy in a wide variety of tasks, such as contact prediction or functional annotation. In late 2022, META AI released a new ESM version, ESM2 (Lin et al. 2023). ESM2 contains 15 billion parameters and performs so extraordinarily that it has been used for protein structure prediction, such as in the over 600 M predictions from the Mgnify database released in the ESMatlas (https://esmatlas.com/). Despite not being a generative model, ESM2 has been used to design *de novo* proteins with remarkable success (Verkuil et al. 2022). Namely, in this work the authors explore designing sequences for a defined backbone by sampling the model via Markov chain Monte Carlo (MCMC) with simulated annealing and concomitantly designing sequence/structure generation by sampling from both distributions (Verkuil et al. 2022).

Decoder-only Transformers have also emerged as very powerful architectures in protein research. Given their autoregressive objective, they are particularly suitable for sequence generation, generating amino acids or k-mers in the N to C-terminus direction. Decoder-only models can be primarily categorized into two types, depending on their conditioning nature. Unconditional language models sample sequences from their learned probability distribution, and by doing so, they generate sequences in unpredictable regions of the protein space. In contrast, conditional models generate sequences conditioned on input data, which may specify a specific property, such as the protein function or target organism. The following section focuses on these types of models.

## 1.1 Unconditioned generative language models

Several recent works have applied decoder-only architectures for unconditional protein design. In 2022, Moffat and colleagues implemented DARK (Moffat et al. 2022), a 110-million-decoder-only transformer capable of designing novel structures, and Ferruz et al. released ProtGPT2 (Ferruz et al. 2022b), a 738 million transformer model based on the GPT-2 architecture. ProtGPT2 showed to generate *de novo* sequences in unexplored regions of the protein space while the sequences exhibited values akin to natural sequences in multiple properties, such as disorder, dynamic properties, pLDDT values, or predicted stability.

RITA is a suite of generative decoder-only models for protein design ranging from 85M to 1.2B parameters, trained on the Uniref100 database for each sequence and its reverse. Besides releasing the models, the authors studied the relationship between model size and downstream performance and observed that performance at predicting protein fitness increased with model size (Hesslow et al. 2022). Nijkamp and colleagues released ProGen2 (Nijkamp et al. 2022), a family of models of up to 6.4B parameters trained on over a billion proteins from genomic, metagenomic, and immune repertoire databases. The models generate sequences predicted to adopt well-folded structures, despite being significantly distant from the current protein space, and are able to predict protein fitness without further training.

One particularly interesting property of protein language models is that they can be used in *zero-shot* or after *fine-tuning* (Radford and Narasimhan 2018) (**Fig. 2d-e**). When used in *zero-shot*, models sample from their entire distribution and hence generate sequences unconditionally. In contrast, *fine-tuning* the model is a process that updates its parameters by learning from a new, narrower dataset, like a protein family. In this case, they will generate new sequences from that group and can be used to augment protein family repertoires (**Fig. 2e**). *Fine-tuning* is possibly the best strategy to add some control over the generation process; alternatively, post-generation filtering can also provide proteins with tailored properties given the fast inference times of these models. In this direction, Ferruz et al. developed a pipeline that combines the synergistic nature of encoder and decoder-only architectures, where ProtGTP2 was used to generate sequences in a high-throughput fashion while ProtT5 annotated their functions (Ferruz et al. 2022b, 2022a).

## 1.2 Conditional language models: tailored protein design

Conditional language models can be trained by coupling the training sequences with control tags, such as their annotated functions or properties; in this way, the model learns a joint sequence-function distribution. One of the most essential works in this direction was the development of the Conditional TRansformer Language (CTRL), an autoregressive model including conditional tags capable of controllably generating text without relying on input sequences (Keskar et al. 2019). These tags, called control codes, allow users to influence genre, topic, or style more specifically - an enormous step towards goal-oriented text generation.

Shortly after CTRL implementation, some of the authors adapted this model to the protein realm, by training on a dataset of 281 million protein sequences (Madani et al. 2020). The model, named ProGen, contains as conditional tags UniparKB Keywords, a vocabulary of ten categories including ‘biological process,’ ‘cellular component,’ or ‘molecular function.’ In total, the conditional tags comprised more than 1,100 terms. ProGen presented ‘perplexities’ representative of high-quality English language models, even on protein families not present in the training set. The generation of random sequences and their Rosetta energy evaluation revealed that the sequences had better scores than random ones. The authors experimentally validated ProGen in the generation of lysozymes after fine-tuning on five different protein families (Madani et al. 2023). The results showed that the generated sequences possess enzymatic activities in the range of natural lysozymes, even in cases with sequence identities as low as 40-50%, and sometimes rival that of a natural hen egg white lysozyme. X-ray characterization of one of the variants showed that it recapitulated the native 3D structure.

In a similar direction, ZymCTRL is a conditional model trained on the *Corpora* of enzymes and their corresponding annotations (Munsamy et al. 2022). Enzymatic sequences are classified depending on their Enzymatic Commission (EC) numbers, identifiers that group sequences that catalyze the same chemical reactions. By training a CTRL-like language model on a set of sequences and their EC classes, ZymCTRL learned a joint distribution of sequence properties and functional annotations and can effectively generate sequences that perform a user-defined enzymatic reaction. An overview of all released protein language models to date is summarized in **Table 1**.

**Table 1.** Summary of released language models ordered by inverse chronological order.

| Encoder-only (embedding) | | | |
|---|---|---|---|
| **Model name** | **Number of parameters** | **Date** | **Reference** |
| **Ankh** | 1.15B | Jan 2023 | (Elnaggar et al. 2023) |
| **ESM2** | 15B | Oct 2022 | (Lin et al. 2023) |
| **DistilProtBert** | 230M | May 2022 | (Geffen et al. 2022) |
| **ProteinLM** | 200M – 3B | Aug 2021 | (Xiao et al. 2021) |
| **ProteinBERT** | 16M | May 2021 | (Brandes et al. 2022) |
| **ESM1** | 43M - 670M | Dec 2020 | (Rives et al. 2021) |
| **PRoBERTa** | 44M | Sep 2020 | (Nambiar et al. 2020) |
| **ProtTrans** | 420M - 11B | Jul 2020 | (Elnaggar et al. 2021) |
| **TAPE** | 38M | Jun 2019 | (Rao et al. 2019) |
| **Decoder-only (generation)** | | | |
| **ZymCTRL** | 762M | Dec 2022 | (Munsamy et al.) |
| **ProGEN2** | 151M- 6.4B | Jun 2022 | (Nijkamp et al. 2022) |
| **RITA** | 85M- 1.2B | May 2022 | (Hesslow et al. 2022) |
| **Tranception** | 700M | May 2022 | (Notin et al. 2022) |
| **ProtGPT2** | 762M | Mar 2022 | (Ferruz et al. 2022b) |
| **DARK** | 110M | Jan 2022 | (Moffat et al. 2022) |
| **ProGEN** | 1.2B | Mar 2020 | (Madani et al. 2020) |

## 2. Non-Transformer generative models

Although transformers have dominated the field of protein sequence generation in recent years, multiple other model architectures have been utilized for sequence generation. While most of these models had the general capability to generate sequences, however, only a few were developed with the specific task of global sequence generation in mind. **Table 2** summarizes the latest non-transformer models with generative capabilities.

**Table 2.** Non-exhaustive list of recently released non-Transformer generative models.

| Model name | Architecture | Features | Reference |
|---|---|---|---|
| **UniRep** | LSTM | Learning of representation | (Alley et al. 2019) |
| **Low-N Unirep** | LSTM + Linear Model | Low-N Protein Fitness Prediction | (Biswas et al. 2021) |
| **ProteinGAN** | GAN | MDH Protein Family | (Repecka et al. 2021) |
| **ProteoGAN** | GAN | Conditional on GO-Terms; Global | (Kucera et al. 2022) |
| **Antibody-GAN** | GAN | Humanoid antibody design | (Amimeur et al. 2020) |
| **FBGAN** | GAN | Peptide Design | (Gupta and Zou 2019) |
| **MSA VAE/AR- VAE** | VAE | Focus on protein family design | (Hawkins-Hooker et al. 2021) |
| **ProteinVAE** | VAE | Generation of synthetic viral vector serotypes | (Lyu et al. 2023) |
| **PepVAE** | VAE | Focus on peptides paired with antimicrobial activity prediction | (Dean et al. 2021) |

LSTMs (Long Short-Term Memory Networks) were the precursors to transformers and were capable of detecting and learning from long-term dependencies as a type of recurrent neural network. UniRep (Alley et al. 2019) is an autoregressive multiplicative LSTM comprising 20.15 million parameters with nine layers that progressively decrease in dimensionality from 1900 to 64 dimensions (mLSTM strategy illustrated in **Fig. 3b**). UniRep's generative ability relies on a given input seed and results mostly in proteins with high identity (>50%) to natural proteins, limiting its potential to explore new areas of the sequence space, but allowing augmenting protein families' repertoires. In the directed evolution section, we will discuss the applicability of UniRep for protein engineering.

ProteinGAN (Repecka et al. 2021) is a generative adversarial network with 45 layers spanning over 60 million trainable parameters. The general strategy of using GANs for sequence design is to first generate sequences in the local/global sequence space by a generator and then let a pre-trained discriminator decide whether the sequence is natural or generated (**Fig. 3a**). This enforces the model to generate sequences in the area of natural clusters. In a first work, the authors assessed the capability of ProteinGAN to generate catalytically active Malate Dehydrogenases (MDH), discovering fully functional MDHs with an average identity to the natural space of 66% (Repecka et al. 2021). In subsequent work, by testing three distinct generative models in the MDH and superoxide dismutase, authors proposed computational metrics for predicting *in vitro* activity that could help in the selection process of active enzymes for experimental characterization (Johnson et al. 2023), demonstrating the potential of these models to generate highly diverse functional proteins with natural-like physical properties.

This approach was recently adopted by ProteoGAN (Kucera et al. 2022), which conditions the generation of *de novo* proteins with labels from the hierarchically sorted Gene Ontology annotations. The method trains with conditional labels and focuses on a global dataset allowing for a general approach to the problem of *de novo* protein design. Amimeur and colleagues presented Antibody-GAN as an approach capable of generating large and diverse libraries of novel antibodies mimicking the somatic response from humans. The method designs variants with improved stability and developability, allowing the control of properties to find suitable therapeutic antibodies (Amimeur et al. 2020). In another application of GANs, FBGAN is a feedback-loop mechanism that generates sequences with a prompted function of the gene product. It was optimized to produce α-helical antimicrobial peptides, showing the opportunity to go from global to local peptide sequence generation through fine-tuning (Gupta and Zou 2019).

Another approach used in the past involves using multiple sequence alignments (MSA) to generate sequences (**Fig 3c**). This method does not result in a complete *de novo* design of proteins but rather an interpolation between the variations in the MSA. The generated proteins are more functional than those from a *de novo* design approach, as they have only minor deviations in the variable amino acid positions from natural proteins. Two different architectures have gained attention in this area: variational autoencoders (VAE) and Potts Hamiltonian maximum entropy models, which are a type of restricted Boltzmann machines. VAEs can recognize higher-order epistasis that is often absent in rational design approaches, a crucial capability for navigating the protein fitness landscape.

Examples of VAEs with the general capability to draw new sequences from their learned distribution are MSA VAE (Hawkins-Hooker et al. 2021) and Deep Sequence (Riesselman et al. 2018). Hawkins-Hooker et al. not only showed in their work that sequence generation with alignment-based VAEs is possible but also that it is possible without aligned input sequences by introducing an autoregressive (AR) component to variational autoencoders. Both the MSA VAE and the AR-VAE models could capture and integrate learned physicochemical properties in their sequence generation. However, the authors emphasize that the models trained on raw sequence inputs miss more long-distance dependencies in the mutational patterns and 3D structure. To experimentally validate these architectures, they trained both models on sequences of the luciferase protein family. By this, they were in both cases able to generate novel proteins with an improved solubility compared to wild-type.

Also related to VAEs but in applications on a more complex protein family with longer sequences, ProteinVAE was developed to generate synthetic viral vector serotypes without epitopes for pre-existing antibodies (Lyu et al. 2023). This VAE used the knowledge learned by a previous model (ProtTrans, Elnaggar et al. 2021) but trained in a natural dataset of adenovirus hexon sequences. ProteinVAE learns intrinsic relationships of long protein sequences and generates diverse proteins with patterns similar to the natural ones. The latent space exploited by ProteinVAE is structured and can be used to facilitate the selection of distant sequences to be tested in the lab.

VAEs also have been used to assist in the generation of bioactive peptides (Das et al. 2018; Dean and Walper 2020; Dean et al. 2021). From them, PepVAE (Dean et al. 2021) is the newest semi-supervised VAE model that designs active novel antimicrobial peptide sequences. By coupling it with antimicrobial activity prediction, PepVAE demonstrates the possibility of VAEs to expand the diversity and functionality of peptides.

Potts Hamiltonian models (**Fig. 3c**) seek to learn the co-variations between any given position in a protein. This covariation matrix correlates and implies, in many cases, the structural contacts inside the given protein. Moreover, these types of models map sequence variations to their corresponding prevalence. They do this by transforming the statistical Potts energy prediction with the Boltzmann distribution. Here, the statistical Potts energy prediction is proportional to the Gibbs free energy and the transformation gives us a probability measure that the protein will occur in that variant. A recent study (McGee et al. 2021) developed a benchmark to examine the generative capacity of probabilistic protein sequence models trained on MSAs and found that Mi3, a Potts Hamiltonian model with only pairwise interaction terms (direct coupling), outperformed various VAEs as well as side-independent models.

Many frameworks to use Potts Models as efficiently as possible were developed and tested. For example, adabmDCA (Muntoni et al. 2021) uses a direct coupling analysis to capture amino acid propensities and other relevant properties, showing promising results in a local sequence generation task that potentially would also be scalable to a global generative capacity. Further development of the algorithm proposed for the adabmDCA framework (Barrat-Charlaix et al. 2021) addresses an important aspect of generative modeling by achieving a parameter reduction to allow reducing the computational cost of generative modeling. In a related work, bmDCA is a model that generates sequences based on pairwise interactions among a large and diverse multiple sequence alignment (Russ et al. 2020). After being tested in the chorismate mutase, artificial enzymes recapitulate the catalytic parameters from natural proteins, demonstrating the potential of evolution-based models to find improved proteins. In summary, MSA-based models are powerful tools to capture pairwise or higher-order mutational variation with high confidence of functional outcomes with comparatively few parameters, but the input MSAs are computationally and data intensive. Lastly, the realm of stable diffusion models for global sequence design has yet to be explored, but the rapid progress of this architecture holds a promising future (Ingraham et al. 2022; Watson et al. 2022).

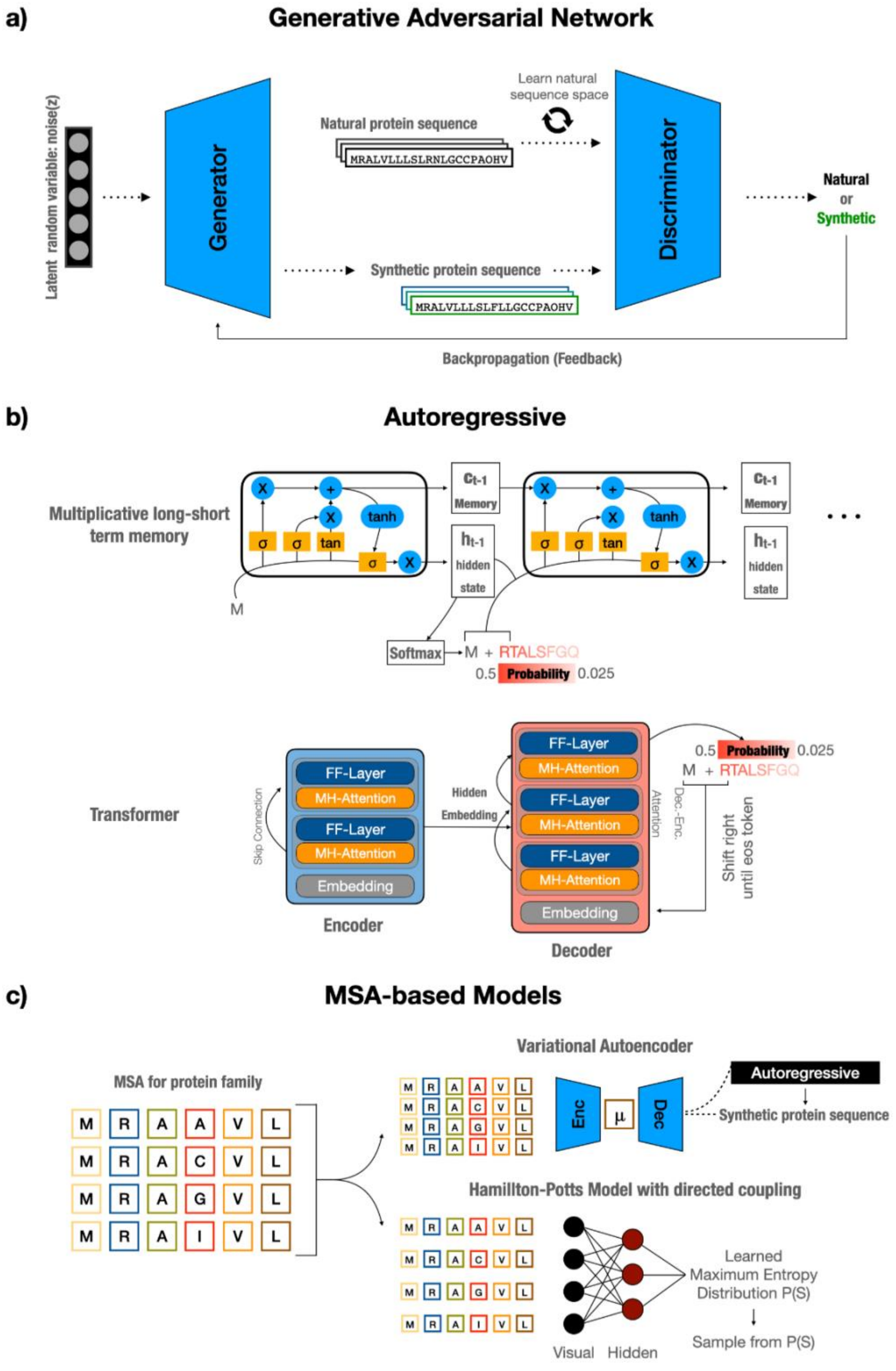

**Figure 3**. **Approaches to global sequence design. a)** GANs as generators from noise with a selection process through the discriminator **b)** mLSTM and Transformer (section 2) as autoregressive models predicting iteratively next token based on the previous ones **c)** Models based on multiple sequence alignments of homologous proteins.

## 3. Assistance in directed evolution techniques and protein design

Directed evolution has been a successful method for creating new biological molecules, such as scaffolds, enzymes, or antibodies, by mimicking the process of natural evolution. It involves creating a population of molecules with a desired property or function, selecting the best ones that perform that function, and then using genotypic and microbiological techniques to create new generations of molecules with even better performance (Kuchner and Arnold 1997) (**Fig. 4**). Directed evolution, as well as generative models, is related to the general concept of *protein sequence space* or *fitness landscapes*, a fundamental idea proposed in early 1970 by Maynard Smith to refer to the vast number of possible meaningful and non-meaningful sequences that can form a protein based on the combination of amino acids, describing the relationship between protein sequences and their functions (Maynard Smith 1970; Ogbunugafor 2020).

This idea suggests that the possible sequences of a protein can be visualized as points on a landscape, with the height of each point representing the functional fitness of the corresponding sequence. Sequences with higher fitness would be located at higher points on the landscape, while less functional sequences would appear at the landscape valleys (**Fig. 4b**). However, it is known that many of the possible protein sequences and evolutionary trajectories will not yield folded and stable three-dimensional structures and, therefore, not any relevant function or fitness (Arnold et al. 2001; Peisajovich and Tawfik 2007; Wagner 2008; Romero and Arnold 2009; Kondrashov and Kondrashov 2015; Starr and Thornton 2016; Wheeler et al. 2016). Consequently, only a reduced subset of sequences will follow an evolutionary trajectory that forms stable and functional proteins. This subset is known as the *protein sequence space of life* and is the set of sequences found in nature or the sequences that can be successfully designed.

Meanwhile, *traditional directed evolution* (TDE) approaches have yielded successful outcomes with high applicability in different fields (MacBeath et al. 1998; Shaner et al. 2004; Aharoni et al. 2005; Fasan et al. 2007; Giger et al. 2013; Kan et al. 2017; Hammer et al. 2017; Zhang et al. 2022) this method has some limitations and challenges (as described in Sellés Vidal et al. 2023), such as i) a time-consuming and expensive process, requiring numerous rounds of mutagenesis-screening-selection; ii) the genetic diversity of a population may be limited, reducing the chance of finding a suitable variant; iii) the outcome can be unpredictable, making difficult to control the final variants; iv) the evolved proteins may not have the desired specificity, leading to off-target effects or reduced efficacy. For these reasons, generative models can significantly help to improve our protein sequence exploration and therefore get faster and more accurate sequence sampling, as has been extensively discussed in a previous review (Wittmann et al. 2021a). Among these models, we can recapitulate some of the most relevant from the last few years.

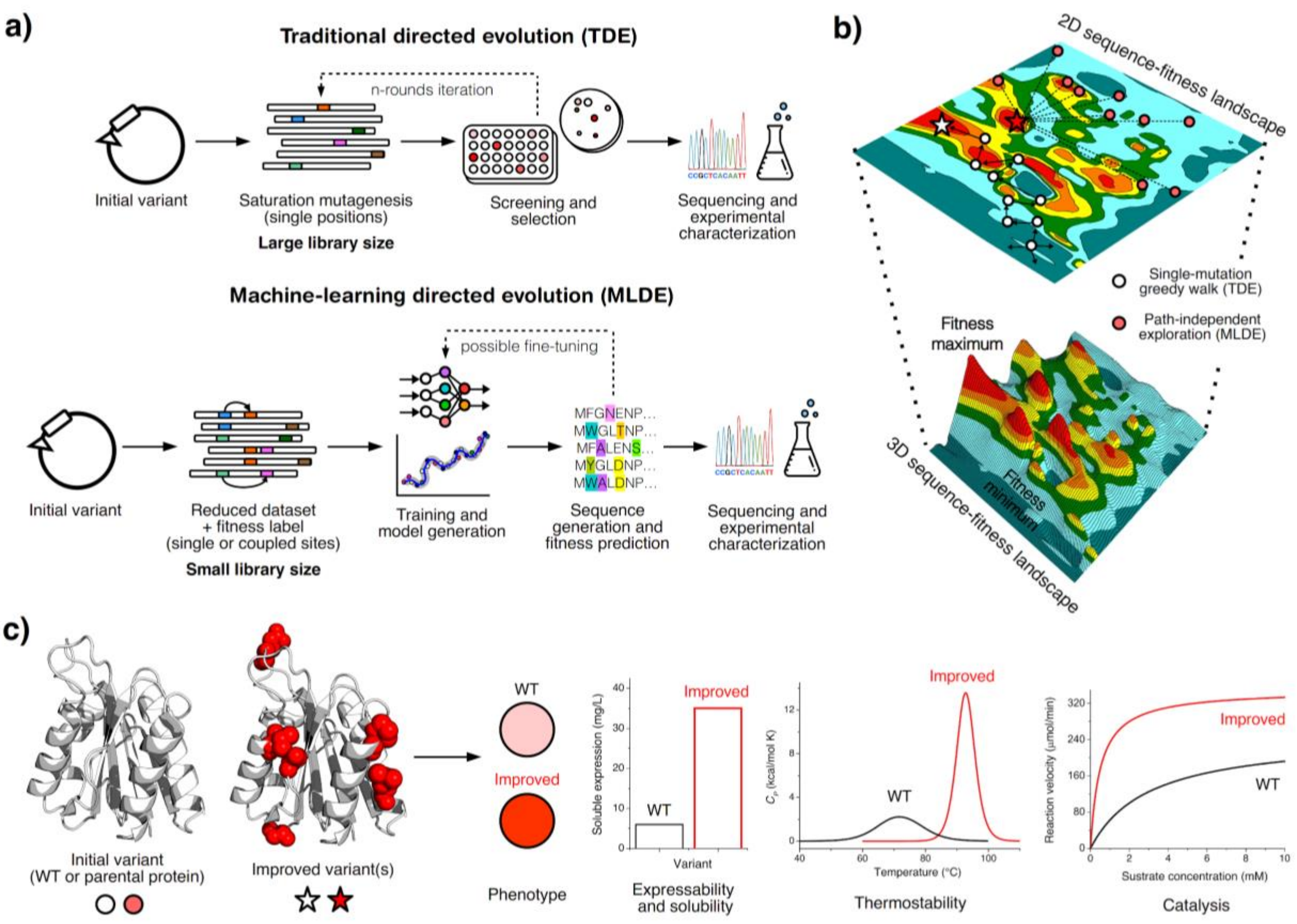

**Figure 4**. **Comparison between traditional directed evolution (TDE) and machine-learning-directed evolution (MLDE) approaches. a & b)** In TDE methods, a large library size is needed to explore broad sequence space and eventually reach variants close to the maximum performance (white star) in the sequence-fitness landscape, a process commonly known as single-mutation greedy walk (solid arrows in panel b). On the other hand, MLDE models use focused training experimental data libraries (small size) with multiple combinatorial positions (labeled or non-labeled) following a path-independent exploration in the fitness landscape (dotted lines in panel b). Then, after training and model generation, it is possible to predict new improved variants (red star) and reach fitness maximums more easily (red zones in the landscape), reducing the necessity of a large experimental characterization. **c)** Newly generated proteins (from both TDE and MLDE) are then analyzed to compare differences among the initial variant and the improved ones on biochemical and biophysical features such as phenotype (e.g., fluorescence, antibiotic resistance, using color-modifying reporter genes, etc.), expressability/solubility, thermostability, or catalytic activity, among others.

Consecutive approaches by the Arnold's group have demonstrated the relevance of *machine learning-assisted directed evolution* (MLDE) to navigate protein epistatic fitness landscapes *in silico*. These approaches, in general, are implemented by training an ML model with a small-size experimentally characterized library, i.e., each combinatorial variant used in the training dataset is labeled with a known experimental fitness (enzymatic activity, stability, function, etc.). Then, the model can be used to predict and explore an entire combinatorial space with co-mutated positions (**Fig. 4**), which enables exploring a larger number of sequences in subsequent analysis rounds (Wittmann et al. 2021b). Using this strategy, a human protein G domain (GB1 binding protein) was used to validate a model fitness landscape by exploring multiple positions simultaneously. Then, to determine if MLDE can find improved variants efficiently, a putative nitric oxide dioxygenase was engineered to evolve its enantiodivergent enzyme activity (Wu et al. 2019). Generated variants demonstrated that sampling *in silico* larger sequence space of a specific protein family favors a rapidly evolving of parent enzymes to selectively perform the desired activity or function, considerably reducing the size of libraries that should be experimentally tested in the lab to obtain similar results.

In an ensuing analysis, MLDE simulations also in the GB1 domain granted to evaluate of the importance of different protein encoding strategies, training procedures, models, and training datasets in reducing the chances of including proteins with extremely low fitness, which minimize the accuracy of these models (Wittmann et al. 2021b). The optimized workflow, by using an informative training set, is able to get up to 81-fold more possibilities to achieve the global fitness maximum on protein engineering tests, demonstrating the potential and out-competing of MLDE in comparison with TDE approaches. Details on the steps and requirements needed to build MLDE models for protein engineering applications are described in a previous review by Yang and colleagues using two case studies that exemplify relevant biological problems: the increase of thermostability of a protein (cytochrome P450) and maximize the productivity of an enzyme (halohydrin dehalogenase) (Yang et al. 2019).

In a related work also training ML models directly on experimental data to explore a full diversity landscape, the adeno-associated virus 2 capsid protein was used as an example system to generate viable engineered capsids varying only in short sequence regions (Bryant et al. 2021). Employing complete, random, and additive sampling strategies to generate experimental libraries, these datasets were used to compare the performance of three model architectures. Convolutional neural networks and recurrent neural networks were the more successful at deep diversification, generating functional variants that are viable for assembly and packing DNA payloads, suggesting that by using small, simple, and unbiased training sets, generative models are able to predict variable, robust, and functional variants.

Coming also in the direction to test the influence of training data on MLDE results, Saito et al. investigated the effect of including or not a known highly active (positive) variant in the initial ML datasets. By performing two separate series and rounds of MLDE in Sortase A, authors found that the independently directed evolution rounds generated 2.2 to 2.5-fold improved variants, and these sequences explored different regions of the fitness landscape depending on the absence or presence of highly positive proteins in the training data, suggesting that the versatility of the improved variants in MLDE can be expanded by the varying the dataset composition (Saito et al. 2021).

In a parallel outcome, aiming to reduce the effort and expenses of the experimental characterization and serving as experimental feedback for future training datasets of MLDE, the evSeq methodology (every variant sequencing) is a new collection of standardized protocols for sequencing variant libraries (Wittmann et al. 2022). Different from traditional directed evolution sequencing, which mostly collects sequences for top-fitness constructs, evSeq gets information from all variants, providing valuable and exhaustive sequence-fitness pairs for supervised model training. With this information, generative models can explore missing regions of sequence landscapes and be used for rapid, low-cost MLDE workflows. Also related to experimental setups that might provide feedback adaptive control to MLDE approaches, PRANCE (phage- and robotics-assisted near-continuous evolution) is a system that comprehensively explores molecular fitness outputs under chemically controlled conditions and optimized consumables and produces variants that can be used as training datasets for further MLDE analysis (DeBenedictis et al. 2022).

Building on the concept of LSTMs, which preceded transformers as described in the previous section, Biswas and colleagues proposed a low-N approach that combines the global knowledge of fundamental functional features learned by UniRep with experimentally characterized proteins of a desired target. Applying this archetype idea, after fine-tuning the UniRep model with two target families, green fluorescent protein and TEM-1 β-lactamase, and characterizing a small number of random mutants of wild-type proteins, it was possible to perform MLDE and create a protein fitness landscape that was used to select candidates with an improved function. From the generated variants, they demonstrated that the candidates are fully-functional and optimized proteins (Biswas et al. 2021). This work showed that unsupervised learning is able to simplify the sequence search by eliminating most of the nonfunctional variants and, at the same time, the possibility of exploiting molecular epistasis and giving prominent applications to directed evolution.

Using a transformer-based architecture, Wu et al. generated a model capable of generating functional signal peptides (SP) that can be used to perform specific protein secretion in living organisms (Wu et al. 2020). This model, trained and validated by using SP-protein pairs from all domains of life, generated novel and diverse SP sequences which, when used on *in vivo* expression of SP-protein pairs for four families, are able to produce not only functional enzymes but also proteins that exhibit activity comparable to natural SPs.

On the side of ancestral sequence reconstruction (ASR), the protocol GRASP (Graphical Representation of Ancestral Sequence Predictions) provides valuable guidance for protein engineering by using ancestral sequences as robust templates for directed evolution (Foley et al. 2022). This work showed how a partial order graph and maximum likelihood inference can perform accurate ASR of large protein families such as glucose-methanol-choline oxidoreductases, cytochromes P450, and dihydroxy/sugar acid dehydratases. In addition, hybrid ancestors can be formed by using insertions and deletions as building blocks, which can also guide further protein engineering purposes.

Tied up to *in silico* analysis of sequence-fitness landscapes, using a latent variable model with nonlinear dependencies in order to capture higher-order and context-dependent constraints, DeepSequence is a probabilistic unsupervised model able to predict the effect of mutations for biological sequence families (Riesselman et al. 2018). This model exhibits high accuracy in a majority of the datasets as well or better among those of site-independent or pairwise-interaction models. Using the β-lactamase family as an example, DeepSequence revealed the latent organization of this sequence family by learning interpretable structure for both macro variation and phylogeny; this information could be applied to other families to explore new regions of sequence space.

In another work to predict new sequences, Lu et al. used a three-dimensional self-supervised convolutional neural network (MutCompute) to identify stabilizing mutations of wild-type and previously engineered PETases. The best variant, FAST-PETase (functional, active, stable, and tolerant PETase), showed higher activity both at 40 and 50 ºC and presented only five mutations in comparison wt PETase. This new enzyme is able to degrade PET fragments embedded in textile fabrics, depolymerize pretreated bottle films, and consume non-physically disrupted melted plastic pucks from entire bottles (Lu et al. 2022).

On the other hand, performing a model-guided sequence generation with two generative language models, ESM-1b (Rives et al. 2021) and ESM-1v (Meier et al. 2021), Hie et al. reported the affinity maturation of clinically relevant human antibodies. After experimentally screening only a small number of variants for each type, these newly antibodies showed a maturation higher for both already highly-mature wild type sequences and unmatured antibodies, representing another example of how MLDE is able to generate quickly diverse and functional proteins (Hie et al. 2022).

Finally, Hsu and colleagues assessed previously published machine-learning methods for protein fitness predictions (many of them discussed in this chapter) and developed a combined approach that uses both evolutionary and assay-labeled data (Hsu et al. 2022). This simple baseline approach makes use of supervised data to enhance, with a linear regression model on site-specific amino acid features, the evolutionary density models such as a hidden Markov model (HMM, Shihab et al. 2013), a Potts model (EVmutation, Hopf, et al. 2017), a VAE (DeepSequence, Riesselman et al. 2018), an LSTM (UniRep, Alley, et al. 2019), and a transformer (ESM-1b, Rives et al. 2021). Augmented models in general had better performance at ranking mutational effects overall, being the augmented DeepSequence VAE the most effective for this purpose.

All these examples demonstrate the versatility and potential of generative models to explore and access remote regions of protein sequence space, not only reaching functionalities similar to natural proteins but also unraveling the latent protein space toward sequences with new properties, efficiencies, and functions.

## Concluding remarks

The field of protein design is undergoing rapid transformation due to the impressive advances in the field of artificial intelligence. In particular, the use of architectures that excel in other areas, such as computer vision and natural language processing, is proving to be highly successful in generating sequences in previously inaccessible regions of the protein space. Several methods, including GANs, VAEs, LSTMs, and Transformer architectures, have shown remarkable results. In this work, we provided an overview of these advances in sequence generation and their potential applications in directed evolution. This progress provides an optimistic outlook, with the potential for designing à-la-carte protein functions and new-to-nature enzymes becoming realistic in the near future.